\documentstyle[aps,twocolumn]{revtex}
\begin{document}
\draft

\title{On the Stability of the Iterated Crank-Nicholson Method
in Numerical Relativity}
\author{Saul A. Teukolsky}
\address{Newman Laboratory, Cornell University, Ithaca, NY 14853}
\date{\today}
\maketitle
\begin{abstract}
The iterated Crank-Nicholson method has become a popular algorithm in
numerical relativity. We show that one
should carry out exactly two iterations {\em and no more}. While the limit
of an infinite number of iterations is the standard Crank-Nicholson
method, it can in fact be worse to do more than two iterations, and
it never helps. We explain how this paradoxical result arises.
\end{abstract}

\pacs{04.25.Dm, 04.20.-q, 04.70.-s}

\section{Introduction}

There is currently a large worldwide effort underway attempting to solve
Einstein's equations numerically for astrophysically interesting
scenarios. The problem is extremely challenging technically.
Among the difficulties is that of finding a finite-difference scheme
that allows a stable time evolution of the system. It is well-known that
{\em implicit} differencing schemes tend to be stable. However, the
difficulty of solving the resulting implicit algebraic equations, especially
in three spatial dimensions, has led most researchers to stay with
explicit methods and their potential instabilities.

Several years ago, Choptuik proposed solving the implicit Crank-Nicholson
scheme by iteration. This would effectively turn it into an explicit
scheme, but hopefully by iterating until some convergence criterion
was met one would preserve the good stability properties of
Crank-Nicholson. The iterated Crank-Nicholson scheme has subsequently
become one of the standard methods used in numerical relativity.

In this note, we point out that when using iterated Crank-Nicholson, one
should do exactly two iterations {\em and no more}. While the limit
of an infinite number of iterations is the implicit Crank-Nicholson
method, it can in fact be worse to do more than two iterations, and
it never helps.

\section{Iterated Crank-Nicholson}

To understand this paradoxical result, consider differencing the simple
advective equation
\begin{equation}\label{advec}
{\partial u \over \partial t}={\partial u \over \partial x}.
\end{equation}
(Many equations in numerical relativity are generalizations of this
form, and the differencing techniques are similar.) A simple first-order
accurate differencing scheme is FTCS (Forward Time Centered Space):
\begin{equation}\label{ftcs}
{u^{n+1}_j-u^n_j \over \Delta t}={u^n_{j+1}-u^n_{j-1} \over 2\Delta x}.
\end{equation}
Here $n$ labels the time levels and $j$ the spatial grid points.

It is a standard textbook result that this scheme is unconditionally
unstable. One sees this with a von Neumann stability analysis: Put
\begin{equation}
u^n_j=\xi^n e^{ikj\Delta x}
\end{equation}
and find that the amplification factor $\xi$ is
\begin{equation}
\xi=1+i\alpha\sin k\Delta x,
\end{equation}
where $\alpha=\Delta t/\Delta x$. Since $|\xi|^2 > 1$ for any choice of
$\alpha$, the method is unconditionally unstable.

Backwards differencing gives a stable scheme:
\begin{equation}\label{backwards}
{u^{n+1}_j-u^n_j \over \Delta t}={u^{n+1}_{j+1}-u^{n+1}_{j-1} \over 2\Delta x},
\end{equation}
for which
\begin{equation}
\xi={1\over 1+i\alpha\sin k\Delta x}.
\end{equation}
Now $|\xi|^2 < 1$ for any choice of
$\alpha$, and so the method is unconditionally stable.

The Crank-Nicholson scheme is a second-order accurate method obtained
by averaging equations (\ref{ftcs}) and (\ref{backwards}). Now one finds
\begin{equation}\label{CNfac}
\xi={1+{1\over 2}i\alpha\sin k\Delta x\over 1-{1\over 2}i\alpha\sin k\Delta x}.
\end{equation}
Since $|\xi|^2 = 1$, the method is stable. It is the presence of the
quantities $u^{n+1}$ on the right hand side of equation (\ref{backwards})
that makes the method implicit.

The first iteration of iterated Crank-Nicholson starts by
calculating an intermediate variable ${}^{(1)}\tilde u$ using equation
(\ref{ftcs}):
\begin{equation}\label{tilde}
{{}^{(1)}\tilde u^{n+1}_j-u^n_j \over \Delta t}
={u^n_{j+1}-u^n_{j-1} \over 2\Delta x}.
\end{equation}
Then another intermediate variable ${}^{(1)}\bar u$ is formed by averaging:
\begin{equation}\label{average}
{}^{(1)}\bar u^{n+1/2}_j=\textstyle{1\over 2}({}^{(1)}\tilde u^{n+1}_j
+ u^n_j).
\end{equation}
Finally the timestep is completed by using equation (\ref{ftcs})
again with $\bar u$ on the right-hand side:
\begin{equation}\label{final}
{u^{n+1}_j-u^n_j \over \Delta t}={{}^{(1)}\bar u^{n+1/2}_{j+1}-
{}^{(1)}\bar u^{n+1/2}_{j-1} \over 2\Delta x}.
\end{equation}
(Iterated Crank-Nicholson can alternatively be implemented by averaging
the right-hand side of equation (\ref{advec}). For linear equations,
this is completely equivalent.)

Iterated Crank-Nicholson with two iterations is carried out in the same way.
After steps (\ref{tilde}) and (\ref{average}), we calculate
\begin{eqnarray}
{{}^{(2)}\tilde u^{n+1}_j-u^n_j \over \Delta t}
& = & {{}^{(1)}\bar u^{n+1/2}_{j+1}-{}^{(1)}\bar u^{n+1/2}_{j-1}
 \over 2\Delta x}, \\
{}^{(2)}\bar u^{n+1/2}_j & = & \textstyle{1\over 2}({}^{(2)}\tilde u^{n+1}_j
+ u^n_j).
\end{eqnarray}
Then the final step is computed analogously to equation (\ref{final}):
\begin{equation}
{u^{n+1}_j-u^n_j \over \Delta t}={{}^{(2)}\bar u^{n+1/2}_{j+1}-
{}^{(2)}\bar u^{n+1/2}_{j-1} \over 2\Delta x}.
\end{equation}
Any number of iterations can be carried out in the same way.

Now consider the stability of these iterated schemes. If we define
$\beta=(\alpha/2)\sin k\Delta x$, and call the FTCS scheme (\ref{ftcs})
the zeroth-order method, then direct calculation shows
that the amplification factors are
\begin{eqnarray}
{}^{(0)}\xi & = & 1 + 2i\beta, \\
{}^{(1)}\xi & = & 1 + 2i\beta -2\beta^2, \\
{}^{(2)}\xi & = & 1 + 2i\beta -2\beta^2 -2i\beta^3, \\
{}^{(3)}\xi & = & 1 + 2i\beta -2\beta^2 -2i\beta^3 +2\beta^4,
\end{eqnarray}
and so on. As one would expect,
these are exactly the same values one gets by expanding
equation (\ref{CNfac}) in powers of $\beta$ and truncating at the
appropriate point.

To check stability, compute $|\xi|^2$ for each of these expressions.
You find an alternating pattern. Levels 0 and 1 are unstable; levels
2 and 3 are stable provided $\beta^2 \leq 1$; levels 4 and 5 are unstable;
levels 6 and 7 are stable provided $\beta^2 \leq 1$; and so on.
Since the stability requirement must hold for all wave numbers $k$,
it translates into $\alpha^2/4 \leq 1$, or $\Delta t \leq 2 \Delta x$.
This is just the Courant condition (the 2 occurs because of the 2 in
eqn.\ [\ref{ftcs}]).

Now we see the resolution of the paradox: While the magnitude of the
amplification factor for iterated Crank-Nicholson does approach 1 as
the number of iterations becomes infinite, the convergence is not
monotonic. The magnitude oscillates above and below 1 with ever
decreasing oscillations. All the cases above 1 are unstable, although
the instability might be very slowly growing for a large number of
iterations.

The {\em accuracy} of the scheme is determined by the truncation error.
This remains second order in $\Delta t$ and $\Delta x$ from the first iteration
on. Doing more iterations changes the stability behavior, but not the accuracy.
Since the smallest number of iterations for which the method is stable
is two, there is no point in carrying out more iterations than this.

Note that there was nothing special about using the advective equation
(\ref{advec}) for this analysis. Similar behavior is found for the
wave equation, written in first-order form
\begin{eqnarray}
{\partial u \over \partial t} & = & v, \\
{\partial v \over \partial t} & = & {\partial^2 u\over \partial x^2},
\end{eqnarray}
with the standard centered difference formula for the second derivative term. 
One recovers the usual Courant condition (without the factor of 2) for
the stable cases.

\section{Acknowledgements}

This work was supported in part by NSF grant PHY 99-00672 and
NASA Grant NAG5-7264 to Cornell University.

\end{document}